\documentclass[twocolumn,amsmath,amssymb,prl]{revtex4}

\usepackage{graphicx}
\usepackage{dcolumn}
\usepackage{bm}

\begin{document}

\title{Spin current induced magnetization oscillations in a paramagnetic disc}

\author{Abraham Slachter}
 \email{A.Slachter@gmail.com}
\author{Bart Jan van Wees}%

\affiliation{Physics of Nanodevices, Zernike Institute for Advanced Materials, University
of Groningen, The Netherlands}

\date{\today}

\begin{abstract}
When electron spins are injected uniformly into a paramagnetic disc, they can precess along the demagnetizing field induced by the resulting magnetic moment. Normally this precession damps out by virtue of the spin relaxation which is present in paramagnetic materials. We propose a new mechanism to excite a steady-state form of this dynamics by injecting a constant spin current into this paramagnetic disc. We show that the rotating magnetic field generated by the eddy currents provide a torque which makes this possible. Unlike the ferromagnetic equivalent, the spin-torque-oscillator, the oscillation frequency is fixed and determined by the dimensions and intrinsic parameters of the paramagnet. The system possesses an intrinsic threshold for spin injection which needs to be overcome before steady-state precession is possible. The additional application of a magnetic field lowers this threshold. We discuss the feasibility of this effect in modern materials. Transient analysis using pump-probe techniques should give insight in the physical processes which accompany this effect.
\end{abstract}

\maketitle

The ability of magnets to inject spin polarized currents into non-magnetic materials has been at the center of research for decades\cite{Zutic}. Not long after the discovery of the Giant Magnetoresistance effect\cite{Baibich}, it was proposed that when spin currents are absorbed in magnetic materials it transfers angular momentum to the magnetization and can excite magnetization dynamics\cite{Slonczewski,Berger}. In the previous decade, it was subsequently demonstrated that this can lead to a tunable magnetization precession\cite{Tsoi,Kiselev,Krivorotov,Kaka,Houssameddine} or it can switch the magnetic state of a nanoscale magnet\cite{Chappert,Albert,Yang}. The former is also known as the spin-torque oscillator.

For the spin-torque oscillator it is typically assumed that all spins in a ferromagnetic disc are strongly coupled together leading to a uniform magnetic moment. The torque which is induced by injecting an out-of-plane spin current into this disc induces oscillations of the magnetic moment. In this article we propose the paramagnetic analogue of this spin-torque oscillator. The concept is shown in Fig. 1. Like the ferromagnetic version, it consists of a disc, but this disc is now paramagnetic. Similarly, a spin current $\vec I_{s}$ is injected which points out-of-plane. Instead of relying on the torque which is provided by the spin-current through exchange interaction with the ferromagnetic moment, the torque is now generated by the magnetic fields from the eddy currents which oppose the magnetization dynamics.

To minimize the energy from a uniform magnetic moment present in a disc a demagnetizing field $\vec B_{d}$ appears. By virtue of the shape of the disc this field points mainly out-of-plane. In the spin-torque oscillator, the ferromagnetic moment precesses around this field such that we can have a tunable magnetization precession. If the injection of spins is uniform throughout the disc and the diffusion fast, a demagnetizing field can still exist. Any possible\footnote{For example thermally excited} in-plane spin accumulation $\vec\mu_{||}$ also precesses around this demagnetizing field. Normally this precession is damped by virtue of spin relaxation.

\begin{figure}[!b!]
\includegraphics[width=8.8cm,keepaspectratio=true]{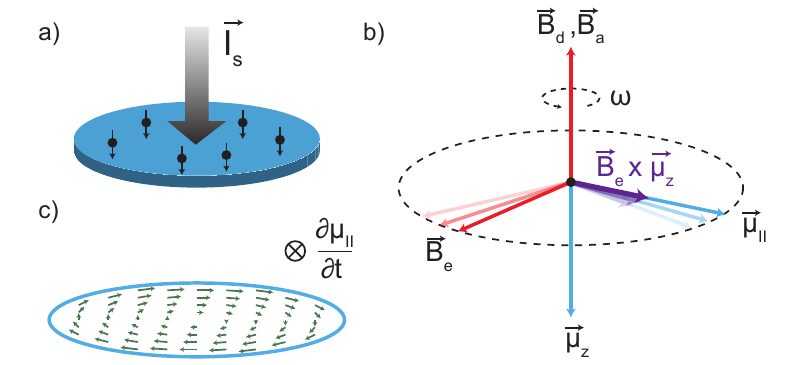}
\caption{\label{fig:1} (Color online) Conceptual diagram. a) An out-of-plane spin current $\vec I_{s}$ is injected into a paramagnetic disc. b) The injected out-of-plane spin accumulation $\vec\mu_{z}$ creates a demagnetization field $\vec B_{d}$ along which any naturally present $\vec\mu_{||}$ precesses. This precession creates a magnetic field $\vec B_{e}$ owing to the eddy currents which can tilt the spin accumulation $\vec\mu_{z}$ in-plane. The effect is enhanced by the application of a static magnetic field B$_{a}$. c) Sketch of the eddy current density in the (spheroid) disc when a changing uniform spin accumulation is present.}
\end{figure}

We propose here a mechanism to create a steady-state precession of this in-plane spin accumulation. When the in-plane spins precess with high frequency, the paramagnetic system tries to oppose the large change in magnetic moment by inducing circulating eddy (or Foucault) currents. These circulating currents produce a magnetic field $\vec B_{e}\sim\omega\mu_{||}$ itself in the opposite direction of the change in magnetic moment, which lags 90 degrees phase with respect to the in-plane spin accumulation. The out-of-plane spin accumulation $\vec\mu_{z}$ can precess around this magnetic field in the direction of the in-plane component. This effectively cancels the relaxation of the in-plane spins leading to a steady-state precession.

Owing to the physical nature of this process, an intrinsic threshold exists. This can be understood as follows. The amount of in-plane spins which relax per unit time is given by $\mu_{||}/\tau$ where $\tau$ is the spin relaxation time. This needs to be compensated by a factor $\sim\vec B_{e}\times\vec\mu_{z}\sim\omega \mu_{z} \mu_{||}$ which also scales with the in-plane spin accumulation $\mu_{||}$. A steady-state precessing spin accumulation $\vec\mu_{||}$ can only exist when the out-of-plane spin accumulation $\mu_{z}$ reaches a certain threshold value $\mu_{z}\sim 1/\tau\omega$ determined by the precession frequency and spin relaxation time. When the out-of-plane spin accumulation $\mu_{z}$ is lower than this threshold the precession simply damps somewhat slower then expected from pure spin relaxation alone. When it exactly matches, the in-plane spin accumulation is stable and precesses at a fixed frequency. By applying an additional magnetic field B$_{a}$ the precession frequency is enhanced which lowers this threshold. In this case the precession frequency $\omega$=g$\mu_{b}/\hbar$ (B$_{d}$+B$_{a}$) is determined by the demagnetization field B$_{d}$ and an applied field B$_{a}$.

In the following, we will calculate the intrinsic threshold in the injected spin current I$_{s}$ of this process. In order to do this, we first calculate the eddy currents which arise in a paramagnetic disc when an uniform change in spin accumulation $\partial\vec\mu_{||}/\partial t$ is present. When a paramagnet is exposed to a changing magnetic field, circulating currents appear in the material to oppose this change. This process is governed by Faraday's law of induction which states $\partial \vec B / \partial t = - \nabla \times \vec E$, where $\partial \vec B / \partial t$ is the change in magnetic field and $\nabla \times \vec E$ is the curl in electric field which is generated. Assuming a steady-state precession of $\vec \mu_{||}$ with frequency $\omega$, the change in magnetic field the disc is exposed to is: $ \partial (\mu_{0}\vec M) / \partial t = C_{e}\omega\mu_{||}$, where we defined a material constant C$_{e}=\frac{1}{4}g\mu_{b}\mu_{0}N_{F}$\cite{Watts,WattsMaser}. Here g is the electron g-factor, $\mu_{b}$ the bohr magneton, $\mu_{0}$ the vacuum permeability and N$_{F}$ the density of states at the fermi level.

\begin{figure}[t]
\includegraphics[width=8.8cm,keepaspectratio=true]{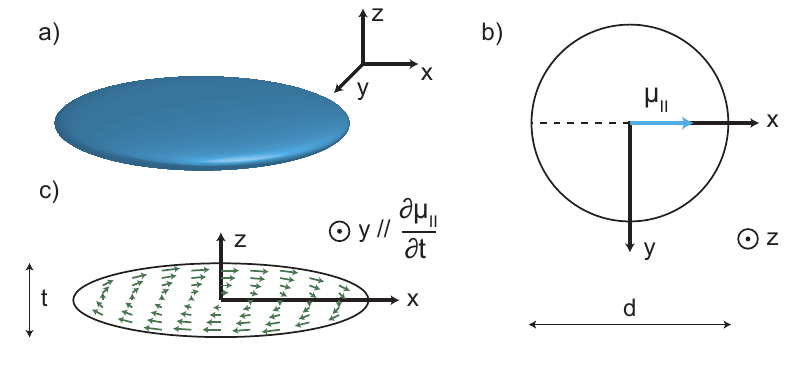}
\caption{\label{fig:2} (Color online) Axis definition eddy fields. a) The disc is approximated by a spheroid with diameter d and thickness t. b) The in-plane spin accumulation $\mu_{||}$ initially points in the x (or $||$) direction while c) $\partial \mu_{||} / \partial t$ points in the -y (or $\perp$) direction. The eddy currents arising in the xz plane are also illustrated.}
\end{figure}

The electric field is determined using charge current continuity along the +x and +z axes (see fig. 2), assuming a uniform $C_{e}\omega\mu_{||}$. The electric field becomes $\vec E = \frac{C_{e}\omega\mu_{||}}{\frac{t}{d}+\frac{d}{t}}\left(\frac{d}{t}z,0,-\frac{t}{d}x\right)$\cite{LandauLifshitzElectrodynamics}. Using the current density $\vec J=\sigma \vec E$, with $\sigma$ the conductivity of the paramagnetic material, we can then obtain the magnetic field from the Biot-Savart law: $\vec B (\vec r_{0}) = \frac{\mu_{0}}{4\pi} \int \frac{\vec J \times (\vec r - \vec r_{0})}{\left|\vec r - \vec r_{0}\right|^{3}}d \vec r$ where the integration is over the volume of the paramagnet. This integral can be evaluated analytically at the center of the spheroid when we assume d$>$10t. The magnetic field at the center of the disc is given by:

\begin{equation}\label{eq:Beddy}
\vec B_{e} = \frac{1}{8}\mu_{0}\sigma C_{e} d t \omega \mu_{||}
\end{equation}

\noindent This field at the center of the conductor scales with the conductivity of the material, the area which is exposed to the changing magnetic field and the precession frequency $\omega$.

In the following, we assume the paramagnetic disc is characterized by a uniform spin accumulation $\vec \mu = \left(\mu_{x},\mu_{y},\mu_{z}\right)$ and feels a uniform eddy field $\vec B_{e}$ as defined by Eq. 1. The dynamics of the spin accumulation $\vec \mu(t)$ can be described by the following equation\cite{Watts,WattsMaser}:

\begin{equation}\label{eq:mu}
\frac{d}{dt}\vec \mu(t) = - \frac{\vec \mu(t)}{\tau}+\vec \omega_{B}\times\vec\mu(t)+\vec I_{s}(t)
\end{equation}

\noindent Where $\hbar\omega_{B}=g\mu_{B}\vec B$ is the Larmor frequency determined by the g-factor and the magnetic field. This equation describes the temporal evolution of spin accumulation. It is determined by spin relaxation $\vec \mu(t)\tau^{-1}$, precession $\vec \omega_{B}\times\vec\mu(t)$ and an external spin injection source I$_{s}$(t) in units of power. We have ignored the spin pumping term previously used\cite{Watts} as well as diffusion. Spin pumping can be shown to be of minor importance since any rotating magnetic fields are only consequences of the effect we describe here and are not directly relevant for the effect itself.

The system of Fig. 1 is solved by injecting a spin accumulation $\vec I_{s}=\mu_{s}\tau^{-1} \hat{z}$ in the disc with $\mu_{s}$ the spin accumulation which would be present in the absence of all magnetic fields. We search for solutions of a steady-state precessing in-plane spin accumulation $\vec \mu=\left(\mu_{||} \cos(\omega t), \mu_{||} \sin(\omega t),\mu_{z}\right)$. We find our solutions in the rotating reference frame\cite{Watts} where the frame rotates with frequency $\omega$ along the z-axis. In this case, the solutions we try to find are static $\vec \mu = \left(\mu_{||},0,\mu_{z}\right)$ such that $\frac{d}{dt}\vec\mu$=0. However, an additional term $\vec\omega\times\vec\mu$ needs to be added to the left side of Eq. 2 with $\vec\omega=\left(0,0,\omega\right)$ defined as the rotation frequency vector of the frame.

The Larmor frequency is determined by the demagnetization field present in the disc as well as the eddy field and is dependent on the spin accumulation $\vec\mu$ present in the disc. The demagnetization field B$_{d} = - \frac{1}{4}g\mu_{B}\mu_{0}N_{F}\bf{N}\cdot$ $\vec \mu$ of a spheroid disc is determined by the magnetic moment present in the disc and the demagnetization tensor $\bf{N}$. The demagnetization tensor for a spheroid disc is diagonal with N$_{\perp}$ on the out-of-plane component and N$_{||}$ on the two in-plane components\cite{LandauLifshitzElectrodynamics} with N$_{\perp}$+2N$_{||}$=1. For a very thin spheroid N$_{\perp}\approx$1.

In the absence of an applied magnetic field, the in-plane components of Eq. 2 can be solved. They provide us the precession frequency and the spin accumulation in the z direction. Using these expressions, the out-of-plane component of equation Eq. 2 provide us the condition for the injected spin accumulation $\mu_{s}$ and the in-plane spin accumulation. We find the following solutions:

\begin{eqnarray}
\mu_{z} &=& -\frac{C_{z}}{N_{F}\sqrt{\sigma\tau d t}} > \mu_{s} \\
\mu_{||} &=& \mu_{z}\sqrt{\frac{\mu_{s}}{\mu_{z}}-1} \\
\omega &=& \frac{C_{\omega}}{\sqrt{\sigma\tau d t}}
\end{eqnarray}

\noindent These three equations are our principal result. The absolute spin current we inject I$_{s}$=$\mu_{s}\tau^{-1}$ needs to be larger then the out-of-plane threshold spin current I$_{thresh}$= $\mu_{z}\tau^{-1}$. The additional spin current I$_{s}$-I$_{thresh}$ which is injected simply increases the size of the in-plane spin accumulation. Here C$_{\omega}=\sqrt{\frac{8\left(N_{\perp}-N_{||}\right)}{\mu_{0}}}$ and C$_{z}=\frac{\sqrt{128}\hbar}{g^{2}\mu_{B}^{2}\mu_{0}^{1.5}\sqrt{N_{\perp}-N_{||}}}$ are two shape-dependent constants. Both the frequency and spin accumulations are inversely dependent on the square root of the conductivity, spin relaxation and area of the spheroid.

The threshold for injected spin accumulation can be lowered by applying an external magnetic field B$_{a}$ in the out-of-plane direction. This way, the precession frequency increases and therefore the eddy current induced magnetic field (eq. 1) as well. When the Larmor frequency of this field $\hbar\omega_{a}=g\mu_{B}B_{a}$ is much larger then the intrinsic precession frequency defined in Eq. 3, the precession frequency is equal to the Larmor frequency $\omega=\omega_{a}$. We can then calculate the new threshold value:

\begin{equation}\label{eq:uzB}
\mu_{z}=-\frac{C_{a}}{N_{F}\sigma\tau d t B_{a}} > \mu_{s}
\end{equation}

\noindent This threshold value for spin injection scales inversely with the applied magnetic field as well as the previously obtained scaling parameters. Here C$_{a}=\frac{32\hbar^{2}}{g^{3}\mu_{B}^{3}\mu_{0}^{2}}$ is another constant. The in-plane spin accumulation remains defined by Eq. 4.

In the following, we wish to determine how feasible this effect is in several known spin systems. The effect becomes feasible in a certain material whenever the amount of spins which can be injected in the system is large enough. This amount is generally determined by a maximum spin injection density J$_{m}$ (spins m$^{-2}$ s$^{-1}$) and the area of the disc. This allows us to  define a feasibility parameter P$_{f}=I_{max}/I_{tresh}$. Here I$_{max}$=J$_{m}$A is the maximum amount of spins per second which can be injected in the disc and I$_{tresh}=\frac{1}{2} N_{F}\mu_{z}/\tau V$ the threshold value (spins s$^{-1}$) for spins which need to be injected before this effect becomes feasible. Here V and A are the volume and area of the disc. Whenever this value is larger then 1, steady-state precession becomes possible and the effect can occur for a given material and spin injection mechanism. In the absence of an external field this parameter becomes:

\begin{equation}\label{eq:Pf}
P_{f}=C_{f}\sqrt{\sigma}\tau^{1.5}J_{m}\sqrt{\frac{d}{t}}
\end{equation}

\noindent This parameter is sensitive to the conductivity, spin relaxation time and the maximum spins you can inject per unit area. Here C$_{f}=\frac{3g^{2}\mu_{B}^{2}\mu_{0}^{1.5}\sqrt{N_{\perp}-N_{||}}}{\sqrt{128}\hbar}$ is yet another shape-dependent constant. When an external magnetic field B$_{a}$ is applied this parameter becomes:

\begin{equation}\label{eq:Pf}
P_{fa}=C_{fa}\sigma\tau^{2}J_{m} d B_{a}
\end{equation}

Where C$_{fa}=\frac{3g^{3}\mu_{B}^{3}\mu_{0}^{2}}{128\hbar^{2}}$ is a constant. In this case, the feasibility of the effect is even more dependent on the material parameters as well as somewhat on the size of the disc and the magnetic field applied. We will use these parameters in the following discussion to judge the feasibility of our effect in different materials. But before we do this, we will first discuss the boundary conditions of our experiment and possible detection mechanisms.

In our analysis, we assumed a uniform spin accumulation in the disc. When we inject spins from the top or bottom of the disc, they need to diffuse in the out-of-plane direction to obtain this situation and feel the demagnetization field. This implies a maximum thickness for the disc in the order of the spin relaxation length $\lambda=\sqrt{D\tau}$ with D the diffusion constant of the material. In addition, we rely on the effect of eddy currents, for which a diffusive transport theory applies. The disc should therefore also have a minimum thickness larger then the mean free path. These requirements give the disc well defined dimensions.

\begin{figure}[t]
\includegraphics[width=8.8cm,keepaspectratio=true]{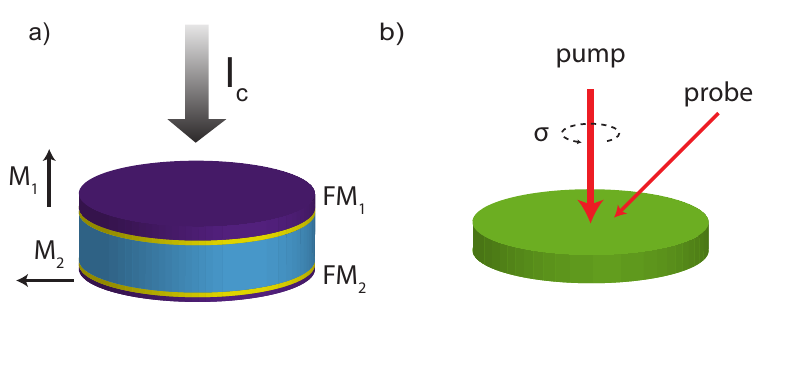}
\caption{\label{fig:2} (Color online) Possible Realization Schemes. a) An out-of-plane spin current can be injected by sending a current from ferromagnet FM$_{1}$ into a paramagnetic metallic disc. The paramagnet is electrically separated by a tunnel barrier. The in-plane spin accumulation can be measured using a second ferromagnet with its magnetization in-plane. b) The effect can be studied by optical spin injection in which a circular polarization for the pump can induce out-of-plane spin polarized carriers. The in-plane component can, for example, be measured using Kerr rotation of a linearly polarized pulse under a small angle.}
\end{figure}

Both the conductivity and the spin relaxation time should be large for the material considered. In practice, the spin relaxation time is limited by the amount of scattering events in a material and often scales inversely with the amount of carriers present in a material. On the other hand, the conductivity scales directly with the amount of carriers present making it difficult to find materials which have both a large conductivity and a high spin relaxation time.

The effect could be measured either electrically or optically as illustrated in Fig. 3. The electrical spin injection scheme and detection scheme can be similar to that used by Houssameddine et al.\cite{Houssameddine} in which there is an out-of-plane polarizer to inject a spin polarized current into the paramagnet while a second ferromagnet is present to analyze the spin accumulation in the paramagnet. The optical spin injection scheme can be realized in a typical pump-probe experiment\cite{Kikkawa} where spins can be injected using circular polarized light with a pump while the in-plane component can be analyzed using Kerr rotation of a linearly polarized probe.

For the electrical spin injection scheme, we first consider aluminum which is known to have a relatively high spin relaxation time of $\approx$100 ps at low temperatures and can have a conductivity as high as 8$\cdot$10$^{7}$ S/m. Let us consider a disc of 1 $\mu$m diameter and 100 nm thickness which satisfies our boundary conditions. The spin injection is limited by the charge current density which can be sent through such materials. In metals, this in the order of 10$^{12}$ A/m$^{2}$. The additional application of a magnetic field of 10T makes this process barely feasible with P$_{af}$=1.05 assuming 100$\%$ spin injection efficiency. This illustrates aluminum is not a very promising material.

Another modern spin-preserving material is graphite which is shown to have a somewhat higher spin relaxation time of $\approx$200 ps\cite{TMaassen}. By doping graphite with an external gate voltage tuned far away from the Dirac point from the single layers of the individual graphene layers, graphite can be metal like. Assuming a uniform $\sigma=3\cdot10^{7}$ S/m and an applied magnetic field of 10T we find a twice higher P$_{af}$. We note that in the out-of-plane direction the conductivity could be less reducing the effective P$_{af}$. This example shows that graphite in its current state is not able to show steady-state magnetization precession.

In semiconductors, the conductivity is strongly reduced. However, the spin relaxation time can be up to 1000 times larger then in metals. Gallium Arsenide is a promising material in which the spin relaxation time can be up to 10-100 ns\cite{Kikkawa} when moderately doped with Si dopants such that the electron density n$\approx$10$^{16}$cm$^{-3}$. The conductivity is then limited to $\approx$5$\cdot$10$^{3}$ S/m. The maximum current density at which spins can be electrically injected is $\approx$10$^{5}$ A/m$^{2}$\cite{Lou}. Optically, this value can be increased up to 100 times, which still makes the maximum spin injection significantly smaller then for ordinary metals. We find that in the best case P$_{fa}$=0.1 with a magnetic field of 10T applied and a disc diameter of 100$\mu$m which shows that our effect is currently not possible in n-GaAs.

Using time resolved pump-probe optical or electrical techniques\cite{Kikkawa,Krivorotov} this effect can be studied directly in the time domain. This allows to study the effect of eddy currents on in-plane precession, even when a steady-state precession is not possible. Indications of this effect should be the observation of a precession frequency determined by the amount of spin accumulation injected in the disc. This should be induced by the demagnetization field even in the absence of an externally applied field. Also, the relaxation time of the in-plane spin accumulation should reduce when more spins are injected.

These examples show, that while it is currently not possible to realize stable precession directly, future improvements in the maximum spin injection in semiconducting systems or higher spin relaxation times in new metallic materials such as graphite will make this effect feasible in the future. In the meantime, transient analysis by means of pump-probe techniques can give insight into the magnitudes of the demagnetization field and the eddy fields which will stabilize this effect.

In conclusion, we proposed the paramagnetic analogue to the previously demonstrated spin-torque-oscillator\cite{Slonczewski,Berger}. The eddy fields which stabilize the precessional motion and the spin accumulations in the disc were calculated using a paramagnetic spin theory\cite{Watts,WattsMaser}. The threshold for the injected spin accumulation which is needed to realize this effect is determined for a general paramagnet. An analysis of this effect in three modern spin-preserving materials, Aluminum, Graphite and n-GaAs shows stable precession is hard to achieve. However, transient analysis using electrical or optical pump-probe technique should give insight in the physical processes which may lead to stable spin precession in future materials.

We acknowledge T.Maassen for critically reading the manuscript. This work was financed by the European EC Contract IST-033749 'DynaMax'.


\end{document}